# Revealing Hidden Spin Polarization in Centrosymmetric van der Waals Materials on Ultrafast Timescales

B. Arnoldi[1], S. L. Zachritz[2], S. Hedwig[1], M. Aeschlimann[1], O.L.A. Monti[2,3*], B. Stadtmüller[1,4#]

1 Department of Physics and Research Center OPTIMAS, Rheinland-Pfälzische Technische Universität Kaiserslautern-Landau, Erwin-Schroedinger-Strasse 46, Kaiserslautern 67663, Germany

2 Department of Chemistry and Biochemistry, University of Arizona, Tucson, Arizona 85721, United States[2]

3 Department of Physics, University of Arizona, Tucson, Arizona 85721, United States[3]

4 Institute of Physics, Johannes Gutenberg University Mainz, Staudingerweg 7, 55128 Mainz, Germany

* monti@arizona.edu
# b.stadtmueller@rptu.de

**Abstract:**

One of the key challenges for spintronic and novel quantum technologies is to achieve active control of the spin angular momentum of electrons in nanoscale materials on ultrafast, femtosecond timescales. While conventional ferromagnetic materials and materials supporting spin texture suffer both from conceptional limitations in miniaturization and in efficiency of optical and electronic manipulation, non-magnetic centrosymmetric layered materials with hidden spin polarization may offer an alternative pathway to manipulate the spin degree of freedom by external stimuli.

Here we demonstrate a novel approach to generate transient spin polarization on a femtosecond timescale in the otherwise spin-unpolarized band structure of the centrosymmetric 2H-stacked group VI transition metal dichalcogenide $WSe_2$. Using ultrafast optical excitation of a fullerene layer grown on top of $WSe_2$, we trigger an ultrafast interlayer electron transfer from the fullerene layer into the $WSe_2$ crystal. The resulting transient charging of the $C_{60}/WSe_2$ interface leads to a substantial interfacial electric field that by means of spin-layer-valley locking ultimately creates ultrafast spin polarization without the need of an external magnetic field. Our findings hence open a novel pathway for optically engineering spin functionalities such as the sub-picosecond generation and manipulation of ultrafast spin currents in 2D heterostructures.

Fundamental to the advance of spintronics and the creation of novel quantum functionalities in solids is the ability to encode, manipulate and store information onto the spin angular momentum of electrons with high efficiency and low volatility [1,2]. Ferromagnetic materials have long been the natural driving target for these efforts. However, their intrinsic limitations regarding miniaturization (i.e., governed by the super-paramagnetic limit) and their susceptibility to external stray fields has triggered a search for non-magnetic materials that can nevertheless support advanced spin functionalities.

One highly promising alternative to ferromagnets are non-magnetic bulk materials with broken symmetries and strong spin-orbit coupling. The combination of both properties results in the spin-splitting of bands in momentum space either through the Dresselhaus effect [3] or the bulk Rashba effect [4–6], and leads to intriguing spin functionalities such as the interconversion of charge and spin [7–9]. Unfortunately, while the resulting spin-momentum locking and associated spin texture does indeed enable controlling the spin degree of freedom, it also limits the type of spin operations that can be realized in such materials: For instance, an unpolarized charge current can only be converted into a transverse spin current by the spin Hall effect [10]. In addition, electric and optical gating, needed for fast operations, can only manipulate the magnitude of the momentum-dependent spin splitting, and both are limited to the picosecond timescale due to the intrinsic buildup time of the photovoltage [11,12]. These fundamental limitations underline the need for new paradigms to tailor and manipulate the spin degrees of freedom, ideally by directly creating and manipulating spin polarization rather than spin texture.

In this regard, the discovery of the so-called *hidden spin polarization* in non-magnetic materials with centrosymmetric crystal symmetry suggests a pathway toward realizing spin manipulation in a much larger class of materials [13–15]. Hidden spin polarizations emerge in centrosymmetric layered structures containing subunits with broken inversion symmetry. Typical examples are, for instance, 2H-stacked group VI transition metal dichalcogenides (TMDs), of which one of the most prominent example 2H-WSe$_2$ is the focus of the present study. A cartoon of the salient features of the spin- and layer-dependent valence band structure of this material is shown in Fig. 1a: It is characterized by spin-split valence bands, localized within each individual layer of the 2H-stacked structure [14,16–18], and whose spin is reversed between the valleys at the high symmetry point K and its time-reversal couple K'. Inversion symmetry of the full bulk unit cell, which contains two layers, leads to an inversion of the valence band spin polarization at each high symmetry point in successive layers, resulting as expected in an overall spin-degenerate bulk band structure. If however the inversion symmetry in otherwise centrosymmetric 2H-WSe$_2$ can be broken between two adjacent layers, e.g. by addressing individual layers differentially, then the emergence of previously hidden spin polarization may be expected, enabling manipulation of spin degrees of freedom without magnetic fields and potentially on ultrafast timescales.

In this work and by using spin- and time-resolved angle-resolved photoemission spectroscopy (ARPES), we overcome this challenge for the first time and demonstrate a new approach to generate transient spin polarization by lifting the spin degeneracy of the bulk band structure at the interface of a C$_{60}$/2H-WSe$_2$ heterostructure. Using ultrafast optical excitation, we are able to generate large interfacial electric fields that ultimately result in ultrafast spin polarization. Conceptionally, this scheme is based on the coupled spin, spin-like valley, and layer pseudospin degrees of freedom that characterize the hidden spin polarization of 2H-WSe$_2$ in the valence band and near the K-points, as expressed by the Hamiltonian [16]:

$$H_v = -\lambda_v \tau_z s_z \sigma_z^v + t_\perp \sigma_x^v \quad (1)$$

Here, the first term describes the coupling between the spin ($s_z$), valley-pseudospin ($\tau_z$), and layer-pseudospin ($\sigma_z^v$) degrees of freedom mediated by spin-orbit coupling $\lambda_v$ (SOC), and the second term describes the coupling of the weak interlayer hopping ($t_\perp$) in WSe$_2$ to the layer pseudospin. Importantly, carrier population in a specific layer represents an interlayer electronic polarization, and hence the layer pseudospin can be considered as an electrical polarizability that can mediate interactions between this spin-like quantity and an external, transient electric field via the Hamiltonian (1) [17–19].

In order to generate the layer-dependent ultrafast electric field, we take advantage of the unique properties of our hybrid organic/inorganic heterostructure by driving interfacial charge-transfer from C$_{60}$ to WSe$_2$ (Fig. 1b). The resulting transient band structure engineering by interfacial electric fields presents the first key step towards ultrafast generation of hole-like spin currents at the interface of TMD bulk materials by fs light excitation, without the need for large external magnetic fields, time-reversal or structural inversion symmetry breaking. Our conclusions are enabled by multi-dimensional photoemission spectroscopy of a C$_{60}$/WSe$_2$ heterostructure with an ultrathin C$_{60}$ layer that allows us to directly access and uncover transient changes of the hidden spin polarization after optical excitation. In this way, we demonstrate that we are able for the first time to trace both the excited state and spin-dependent band structure dynamics at this hybrid heterointerface on the fs timescale.

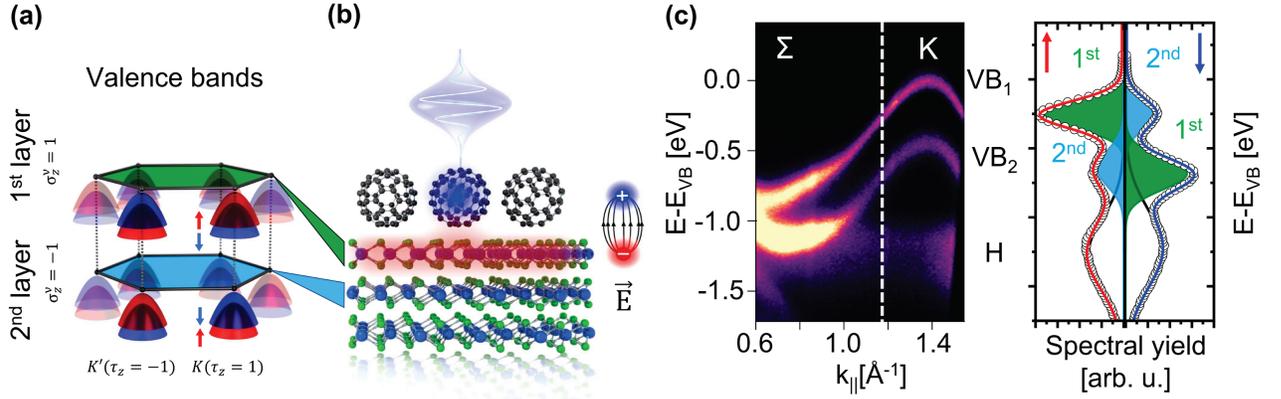

**Fig. 1: Electronic valence band structure of the $C_{60}$/$WSe_2$ heterostructure.**
(a) Sketch of the local layer- and spin-dependent band structure of the two non-interacting $WSe_2$ layers of the bulk unit cell in which the spin polarization vanishes at every point in the Brillouin zone. (b) Illustration of the optical manipulation scheme for uncovering the hidden spin polarization of $WSe_2$. An ultrashort 3.2 eV laser pulse resonantly excites the ultrathin $C_{60}$ layer grown on top of $WSe_2$ leading to an ultrafast electron transfer into the first $WSe_2$ layer and to a transient E-field across the $C_{60}$/$WSe_2$ interface. (c) Energy vs. momentum photoemission map of the $C_{60}$/$WSe_2$ heterostructure along the Σ-K-direction (He I$_\alpha$ radiation). It shows the spin split $WSe_2$ valence bands with their hole-like dispersion ($VB_1$, $VB_2$) and the dispersion-less HOMO (H) of $C_{60}$. The right side of (c) shows the spin-resolved photoemission yield (out-of-plane spin component) of the valence band structure obtained at a selected electron momentum (see white dashed line). The red and blue curves represent the fit to the spin-up and spin-down spectrum, respectively. The contributions of the first and second layer valence bands to the spectral yield are fitted and illustrated as green and blue Gaussian curves underneath the spectra.

Our sample consists of an in situ prepared surface of a 2H-$WSe_2$ bulk crystal covered with approx. 0.8 ML of $C_{60}$ (see method section for more details). The energy level alignment of the valence band structure of the $C_{60}$/$WSe_2$ heterostructure prior to ultrafast excitation can be deduced from the momentum-resolved photoemission map in Fig. 1c, recorded along the Γ-Σ-K high symmetry direction and shown in the vicinity of the K-point. The spin-split valence bands of $WSe_2$ appear as hole-like parabolic features at the K-point with an energy splitting of 450 meV, similar to the bare $WSe_2$ surface (see SI and Ref. [14]), indicative of physisorptive interactions at the $C_{60}$/$WSe_2$ interface. The non-dispersive feature at $E$-$E_{VB}$ = 1.3 eV is attributed to the $C_{60}$ valence state, i.e. the highest occupied molecular orbital (HOMO), and reflects the large ionization energy of $C_{60}$ [20,21]. The spin-resolved photoemission yield of the valence band structure is shown on the right for a selected electron momentum (indicated by a white vertical line in Fig. 1c). The red curve corresponds to the yield of spin-up electrons (out-of-plane spin direction), and the blue curve to the yield of spin-down electrons. The $C_{60}$ HOMO (H) is not spin-polarized, as expected for molecular films on non-magnetic surfaces. In contrast, we find strong spin polarization for both SOC-split $WSe_2$ valence bands ($VB_1$ and $VB_2$). Though bulk $WSe_2$ does not support a spin-split density of states [13], the layer-dependent hidden spin polarization of inversion-symmetric bulk $WSe_2$ is made apparent by the extreme surface sensitivity of the photoemission process [14]. In ARPES, we primarily probe the top $WSe_2$ layer, and hence the photoemission yield at $VB_1$ carries mostly spin-up electrons (green curve/area) near K, while the smaller signal in the spin-down channel (blue curve/area) stems from the inverted spin-polarization of $VB_1$ of the second layer. The opposite is true for the lower valence band $VB_2$. Key to these observations is the fact that our spin- and momentum-resolved photoemission experiment also provides layer-sensitivity. This allows us to disentangle ultrafast momentum-, spin- and layer-dependent band structure changes in the $C_{60}/WSe_2$ heterostructure following optical excitation.

Optical excitation of the heterostructure with 3.2 eV sub-50 fs pulses creates a transient electric field across the $C_{60}/WSe_2$ interface: At this energy, the $CT_2$ state of $C_{60}$ is excited (see energy level alignment diagram in Fig. 2a), supported by previous studies of the two materials [19,22,23]. This state is associated with the formation of intermolecular charge transfer excitons. Using time- and momentum-resolved photoemission, we follow the ultrafast charge-carrier dynamics subsequent to excitation at 3.2 eV. Example energy vs. momentum cuts from these data are shown in Fig. 2b at three characteristic time delays, showing clearly interfacial charge transfer from $C_{60}$ into the $WSe_2$ layer followed by scattering in the $WSe_2$ CB. The experimental data in the excited states are plotted as difference maps (accumulation of spectral yield shown in red, depletion in blue), while the transient changes in the valence band region are shown as intensity maps. Upon excitation ($t = 0$ fs), a broad distribution in momentum space is created at the energy of the $CT_2$ state, accompanied by an instantaneous intensity reduction of the HOMO feature. These optically induced modifications of the electron and hole population coincide with transient linewidth broadening of the interfacial valence band structure which was recently identified as a spectroscopic signature of charge-transfer excitons in molecular films [22,23]. Crucially, we only observe a marginal depletion of the $WSe_2$ valence states (see SI). This proves that the formation of charge-transfer excitons in the $C_{60}$ layer is indeed the dominant optical excitation path and that direct excitation of $WSe_2$ [24] does not play a dominant role here.

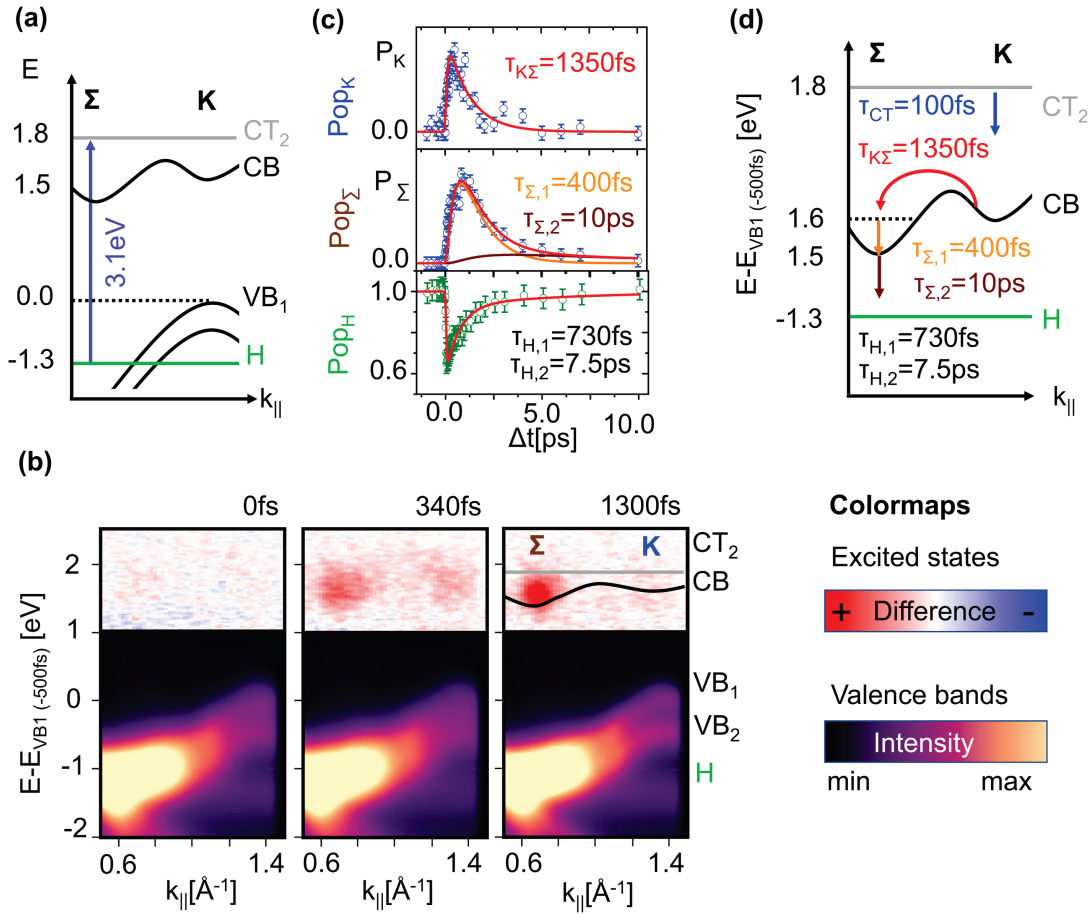

**Fig. 2: Ultrafast electron and hole dynamics**
(a) Energy level diagram of the electronic band structures of the $C_{60}$/$WSe_2$ heterostructure. The blue arrow indicates the dominant optical transition of the 3.2 eV excitation. (b) Energy vs. momentum photoemission maps at selected pump-probe delays obtained with linearly polarized pump pulses (fluence $F$ = 0.5 mJ/cm$^2$). The excited state region ($E-E_{VB} > 0$ eV) is shown as a difference map, the valence band region as an electron intensity map (see colormaps). The energy and momentum positions of the molecular $CT_2$ state and the $WSe_2$ valence band are superimposed onto the experimental data (1300 fs) as gray and black curve. (c) Temporal evolution of the $WSe_2$ excited state and $C_{60}$ HOMO intensity evolution. The solid lines superimposed onto the population dynamics at the K- and Σ-point ($Pop_K$ and $Pop_\Sigma$) were obtained by a rate equation model. The key scattering processes of this model are illustrated in (d) together with the scattering times of the best fit to the data. The temporal evolution of the HOMO is modelled with a double exponential fit function.

This broad electron distribution ($t = 0$ fs) evolves to populating the K- and $\Sigma$-valley of the $WSe_2$ conduction band ($t = 340$ fs), clearly indicating ultrafast electron-transfer from $C_{60}$ to $WSe_2$. Subsequently, phonon-mediated intervalley scattering redistributes carriers from the K-valley into the $\Sigma$-valley of $WSe_2$ ($t = 1300$ fs), a well-known process in TMDs [19,25,26].

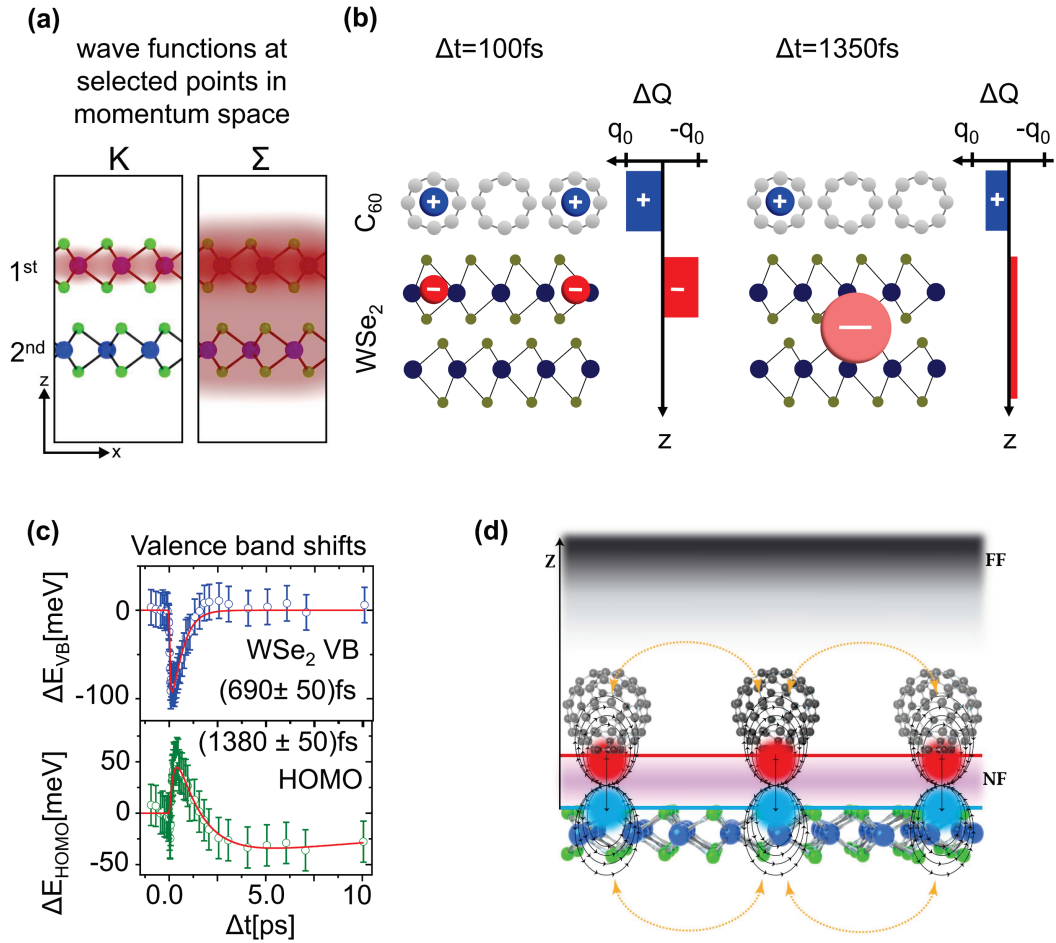

**Fig. 3: Charge separation, interfacial E-field, and transient changes in the energy level alignment.**
(a) Sketch of the real space electron densities (red shaded areas) of the wave functions at the K- to the Σ-valley of the WSe$_2$ conduction band (adapted from Bertoni et al. [19]). (b) Illustration of the charge separation process at the C$_{60}$/WSe$_2$ interface. After the ultrafast electron transfer from the C$_{60}$ CT$_2$ state into the WSe$_2$ K-valley, the electrons are confined to the first WSe$_2$ layer. Only the intervalley scattering form the K- into the Σ-valley leads to a delocalization of the electrons in WSe$_2$. (c) Temporal evolution of the valence band shifts of the WSe$_2$ (VB) and the C$_{60}$ (HOMO) valence states. The dynamics of the energy shifts was analyzed with exponential functions. (d) Electrostatic model estimating the transient valence band shifts.

To gain a more quantitative understanding of the interlayer and intervalley scattering processes at the C$_{60}$/WSe$_2$ interface, we model the changes in electron and hole populations using a rate-equation model (see SI). We extract the transient electron population at the K- and Σ-valley of the WSe$_2$ conduction band and the hole population of the C$_{60}$ HOMO by analysis of the ARPES data (see SI). The resulting traces for all three features are shown in Fig. 2c. Note that no clear population signal could be extracted for the broad $CT_2$ feature due to its large energetic overlap with the $WSe_2$ conduction band. The rate equation model considers the excited state scattering pathways illustrated in Fig. 2d. We assume initial population of the $CT_2$ state by the laser pulse, followed by electron-transfer processes from the $CT_2$ state into the K-valley as well as intervalley scattering from the K- into the $\Sigma$-valley. Additional electron-transfer processes, such as a direct charge transfer from the $CT_2$ level into the $\Sigma$-valley, were not needed to obtain a satisfactory fit of the population dynamics. This yields an interfacial electron transfer time of $\tau_{CT}$ = 100±50 fs, an intervalley scattering time of $\tau_{K\Sigma}$ = 1350±50 fs as well as a depopulation time of the electrons in the $\Sigma$-valley of about 10 ps. The intervalley scattering time $\tau_{K\Sigma}$ is approx. 20 times larger than previously reported for bare $WSe_2$ [19]. We believe that this can be attributed to a sample temperature of approx. 40 K, much lower than in the previous report and causing significantly reduced electron-phonon scattering [27]. A detailed analysis of the hole population dynamics in Fig. 2c reveals a clear persistence of holes in the $C_{60}$ layer. Detailed analysis using exponential fit functions (discussed in Ref. [22]) shows instantaneous depletion of the $C_{60}$ HOMO within our experimental resolution, as expected from resonant excitation, followed by decay of the hole population in a two-step process with a fast recovery time constant of 730±50 fs and a significantly slower second time constant of approx. 7.5 ps.

Summarized in Figs. 3a and b, the key processes involve ultrafast interfacial charge-transfer from $C_{60}$ to $WSe_2$, resulting in a hole located on $C_{60}$ and an electron in the K-valley of the top layer of $WSe_2$. This is followed by intervalley scattering to $\Sigma$, whose electron density spans both the first and second layer of $WSe_2$ [19]. This charge separation between $C_{60}$ and $WSe_2$ establishes a strong and transient interfacial electric field along the surface normal. As we show below, this field is ultimately responsible for revealing the hidden broken inversion-symmetry in WSe$_2$. Eventually, the photoexcited electrons delocalize into the bulk WSe$_2$ crystal and the interfacial electric field decays.

We next discuss the influence of the transient electric field on the interfacial energy level alignment. As can be seen in Fig. 3c, both WSe$_2$ valence band and the C$_{60}$ HOMO experience transient energy shifts, extracted from the energy distribution curves at K (see SI). We attribute these to transient Stark shifts caused by the large electric field built up by the interfacial charge-transfer [28–31]. As expected for such Stark shifts, the sign of the shifts differs for the hole-enriched C$_{60}$ HOMO band and the valence band of the electron-enriched WSe$_2$: A simple electrostatic model of this interface [32] (see SI for details) that considers a distribution of holes residing on C$_{60}$ and electrons residing in WSe$_2$ as illustrated in Fig. 3d shows how the WSe$_2$ valence band and the C$_{60}$ HOMO features are expected to exhibit opposite energy shifts with different magnitudes and time evolution. Scattering of the electron into the Σ valley is accompanied by delocalization into the 2$^{nd}$ layer, weakening the interfacial field. Both these observations and their explanation are also consistent with transient energetic shifts observed in recent experiments on the interface of WS$_2$ on graphene [31]. We conclude therefore that the observed dynamics are indeed driven by a layer-dependent interfacial Stark effect.

The interfacial electric fields hold the key to establishing layer-dependent transient spin polarization: The electric field experienced by the first and second WSe$_2$ layer differs and is coupled to the layer pseudospin, and since each layer is spin-valley-layer locked, the spin degeneracy of the bulk crystal is locally lifted in the first two layers. We investigate transient changes of the valence band spin polarization by monitoring the time-evolution of the spin- and layer-dependent WSe$_2$ valence band structure in the vicinity of the K-point. The corresponding spin-dependent photoemission yield is shown in Fig. 3a for three characteristic instances in time, namely before the optical excitation ($t = -500$ fs), coincident with the optical excitation and initial formation of the charge transfer excitons in the C$_{60}$ layer ($t = 0$ fs), and in the presence of the charge-separated state at the C$_{60}$/WSe$_2$ interface ($t = 950$ fs). The spectra in the left column correspond to spin-up electrons, and those in the right column to spin-down. Fitting these spin- and time-resolved ARPES spectra reveals differential shifts of the valence bands VB$_1$ and VB$_2$ for the first (green Gaussian curve) and second (blue Gaussian curve) WSe$_2$ layer, summarized in Fig. 3b.

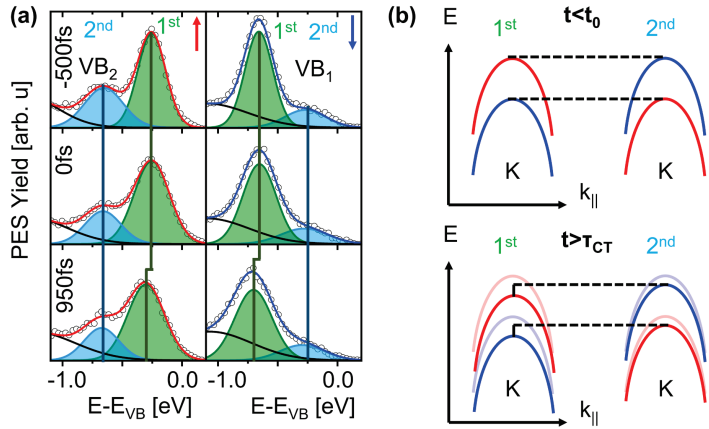

**Fig. 4: Ultrafast changes of the hidden spin polarization of the WSe$_2$ bulk band structure.** (a) Time- and spin-resolved photoemission yield (out-of-plane spin component) of the valence band structure (see white dashed line) at three characteristic time delays. The data were recorded at the same electron momentum as the static data in Fig. 1c. The red and blue curve represents the fit to the spin-up and spin-down spectrum, respectively. The contributions of the first and second layer valence bands to the spectral yield are fitted and illustrated as green and blue Gaussian curves underneath the spectra. The vertical solid lines indicate the significantly larger shift of the valence band of the first WSe$_2$ layer compared to the second layer. (b) energy level diagram illustrating the ultrafast changes of the layer- and spin-dependent WSe$_2$ valence band structure after optical excitation with 3.2 eV photons.

Initially, upon optical excitation ($t = 0$ fs), no changes in the energy level alignment and the spin polarization are observed for any of the WSe$_2$ valence bands, and all spectral changes can be attributed to an instantaneous inhomogeneous linewidth broadening caused by the formation of the C$_{60}$-based CT$_2$ exciton. However, once interlayer charge transfer takes place and the charge-separated state is created ($t = 950$ fs), the WSe$_2$ valence bands shift. In both spin channels, the valence bands of the first layer (green Gaussian curves) transiently shift rigidly towards larger binding energies by 50±20 meV, while the valence bands of the second layer reveal only a minor shift of 20±20 meV. Thus, interlayer charge transfer at the C$_{60}$/WSe$_2$ heterointerface modifies the band structure in a layer-dependent fashion. Crucially, the electric field gradient within the first two WSe$_2$ layers is strong enough to lead to a sizeable relative shift of the spin-polarized bands of the first vs. the second WSe$_2$ layer, thus creating a transient ferromagnetic-like spin polarization in the WSe$_2$ valence bands by revealing the hidden spin polarization in the surface region of the bulk crystal on ultrafast timescales.

In conclusion, our work has demonstrated a novel approach to transiently engineer the spin-polarized valence band structure in the otherwise spin-degenerate layered bulk material 2H-WSe$_2$. Specifically, the ultrafast electron transfer from an optically excited C$_{60}$ layer grown on top of WSe$_2$ leads to a layer-dependent shift of the spin-valley-layer locked WSe$_2$ valence band structure that ultimately reveals the hidden spin polarization of the system on a femtosecond timescale. Our optical manipulation scheme for generating a ferromagnetic-like spin polarization in the valence band without an external magnetic field constitutes not only an avenue for optically engineering new spin functionalities, such as the generation of spin-polarized hole currents in WSe$_2$, on ultrafast, sub-picosecond timescales, but also opens the intriguing possibility for exploiting and manipulating the orbital degree of freedom of layered TMDs thus paving the way for pushing the emergent field of orbitronics [33] towards ultrafast timescales.

**Acknowledegements:** The experimental work was funded by the Deutsche Forschungsgemeinschaft (DFG, German Research Foundation) - TRR 173 - 268565370 Spin+X: spin in its collective environment (Projects A02) and by the Air Force Office of Scientific Research under award FA9550-21-1-0219. B. Stadtmüller acknowledges financial support from the Dynamics and Topology Center funded by the State of Rhineland Palatinate.

**Competing interest declaration:** The authors declare no competing interests.

**Data availability:** The data supporting the findings of this study are available from the corresponding author upon request.

**Methods:**

**Sample preparation:** All sample preparation and measurement steps were performed under ultrahigh vacuum (UHV) conditions. The WSe$_2$ single crystals were obtained from HQ graphene and cleaved prior to the experiments resulting in a clean and flat surface. C$_{60}$ molecules were evaporated onto the surface at a pressure <10$^{-8}$ mbar using a Knudsen-type evaporation source (Kentax GmbH). The molecular flux was calibrated using a quartz crystal oscillator gauge and the molecular coverage was estimated using the integrated intensity signal of the highest occupied molecular orbital of C$_{60}$ as a reference.

**Spin- and time-resolved angle-resolved photoemission spectroscopy (ARPES):** The multidimensional photoemission experiments were conducted with a hemispherical analyzer (SPECS Phoibos 150) that is equipped with both a CCD detector system and the commercial spin detector (Focus FERRUM [34]) that is mounted in a 90° geometry after the hemispherical analyzer's exit slit plane. All spin-resolved photoemission data were recorded for the out-of-plane spin component, i.e., the spin component parallel to the optical axis of the analyzer lens optics. The spin sensitivity or Sherman function (S) of this very-low-energy electron diffraction (VLEED) detector was determined to be 0.29 for the out-of-plane spin component.

As excitation sources, we used the monochromatic He I$_\alpha$ radiation (21.2 eV, Scienta VUV5k) of a high-flux He discharge source as well as a pulsed femtosecond extreme ultraviolet (fs-XUV) light source. The fs-XUV radiation (22.2 eV, horizontal (p) polarization) was obtained by high harmonic generation (HHG) using the second harmonic (390 nm) of a titanium sapphire laser amplifier system (repetition rate 10 kHz, pulse duration < 40 fs) to drive the HHG process [35].

The optical excitation of the organic material was also performed with the second harmonic of the amplifier system (3.17 ± 0.04 eV, bandwidth 80 meV, horizontal (p) polarization). Prior to each time-resolved experiment, the spatial overlap between the pump and the probe pulse was optimized directly on the sample plate, which was placed at the focus position of the analyzer. The spatial overlap was actively stabilized during the experiment to correct for spatial drift of the pump and probe beams. This is achieved by constantly monitoring the beam position of the fundamental laser beam at two well-defined positions in the laser beamline using two CCD cameras. Any lateral draft of the laser beam is compensated by two motorized mirrors installed in the beamline. All time-resolved photoemission experiments were conducted in normal incidence geometry and an emission angle of approx. 45°. A detailed description of the data analysis procedure can be found in the Supplementary Information.

# Revealing Hidden Spin Polarization in Centrosymmetric van der Waals Materials on Ultrafast Timescales

## *Supplementary Information*

B. Arnoldi[1], S. L. Zachritz[2], S. Hedwig[1], M. Aeschlimann[1], O.L.A. Monti[2,3*], B. Stadtmüller[1,4#]

1 Department of Physics and Research Center OPTIMAS, Rheinland-Pfälzische Technische Universität Kaiserslautern-Landau, Erwin-Schroedinger-Strasse 46, Kaiserslautern 67663, Germany

2 Department of Chemistry and Biochemistry, University of Arizona, Tucson, Arizona 85721, United States[2]

3 Department of Physics, University of Arizona, Tucson, Arizona 85721, United States[3]

4 Institute of Physics, Johannes Gutenberg University Mainz, Staudingerweg 7, 55128 Mainz, Germany

* monti@arizona.edu
# b.stadtmueller@rptu.de

**Supplementary Figures**

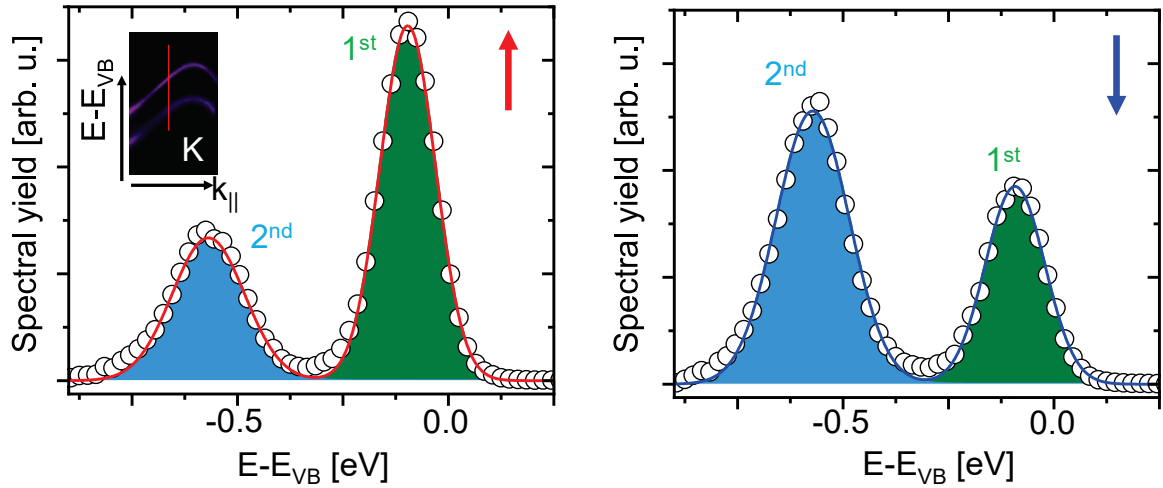

**Fig. S1: Spin-polarized photoemission yield of the 2H-WSe$_2$ valence band structure.**
Spin-resolved photoemission yield (out-of-plane spin component) of the valence band structure obtained at a selected electron momentum (see red line in the inset of the left panel). The red and blue curves represent the fit to the spin-up and spin-down spectra, respectively. The contributions of the first and second layer valence bands to the spectral yield are fitted and plotted as green and blue Gaussian curves below the spectra. Our data analysis reveals a splitting of both valence bands of (470 ± 20) meV which is in agreement with previous photoemission studies [1].

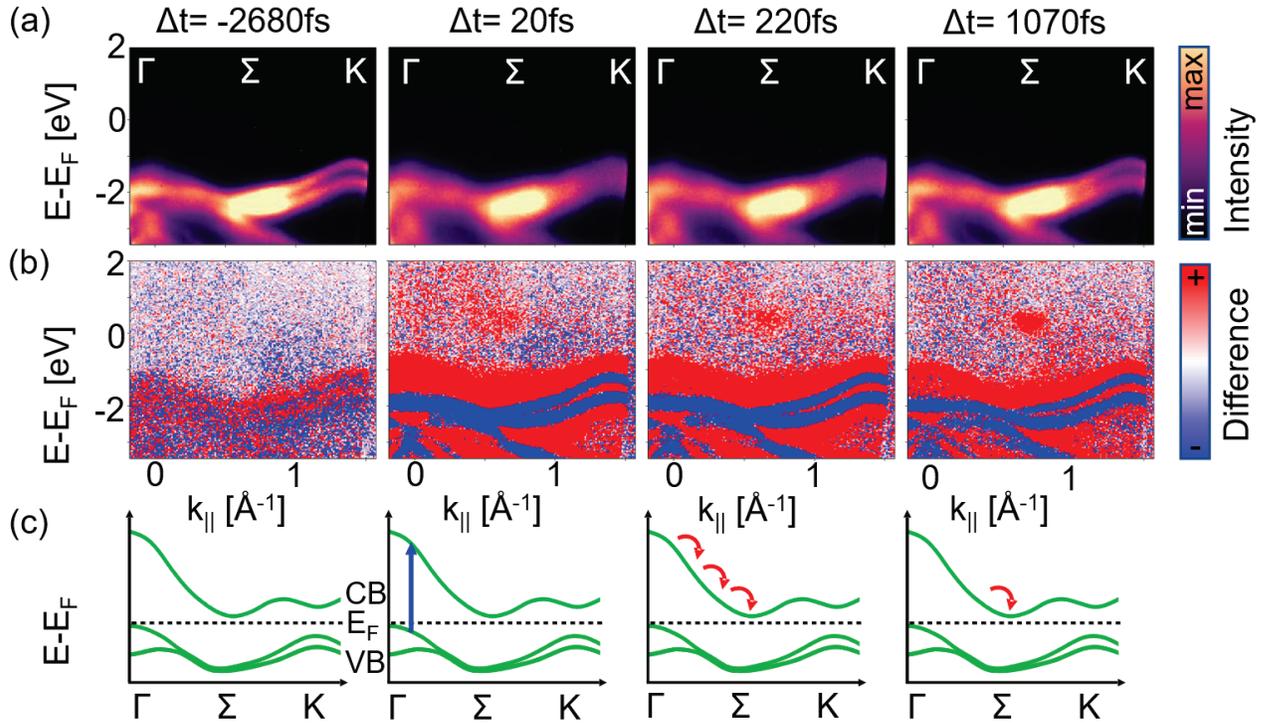

**Fig. S2: Ultrafast dynamics of the bare WSe$_2$ centrosymmetric bulk crystal.**
(a) Example energy vs. momentum intensity maps at selected time delays after optical excitation of a freshly cleaved WSe$_2$ bulk crystal with 3.2 eV photons (fluence $F = 0.5$ mJ/cm$^2$). The photoemission data around the Γ point were obtained with an angle of incidence of the pump and probe beam of 45°, the data at the K-point in normal incidence geometry. The WSe$_2$ valence bands show a substantial transient linewidth broadening, previously reported by M. Puppin et al. [2], which was attributed to an increased phase space for electron-electron scattering due to the depopulation of the WSe$_2$ valence bands. (b) Intensity difference maps of the same energy-momentum maps shown in (a). The difference maps were calculated by subtracting an intensity map before optical excitation (averaged over several time steps) from the intensity map at a selected time delay. In these difference maps, red areas indicate accumulation of spectral intensity, while blue areas indicate loss of spectral intensity. Changes in the excited states can be directly related to the population dynamics of the optically excited carriers. For the bare WSe$_2$ crystal, we observe an instantaneous population of the conduction band followed by intraband scattering of the carriers towards the Σ-valley, as shown in the energy level diagram in (c). In particular, we do not observe a strong population of the K-valley of the conduction band after optical excitation with 3.2 eV photons. This is significantly different from the situation of the C$_{60}$/WSe$_2$ heterostructure discussed in the main manuscript as well as in Fig. S3. Our results are thus in qualitative agreement with the recent work of Puppin et al. for a bare 2H-WSe$_2$ bulk crystal [2,3].

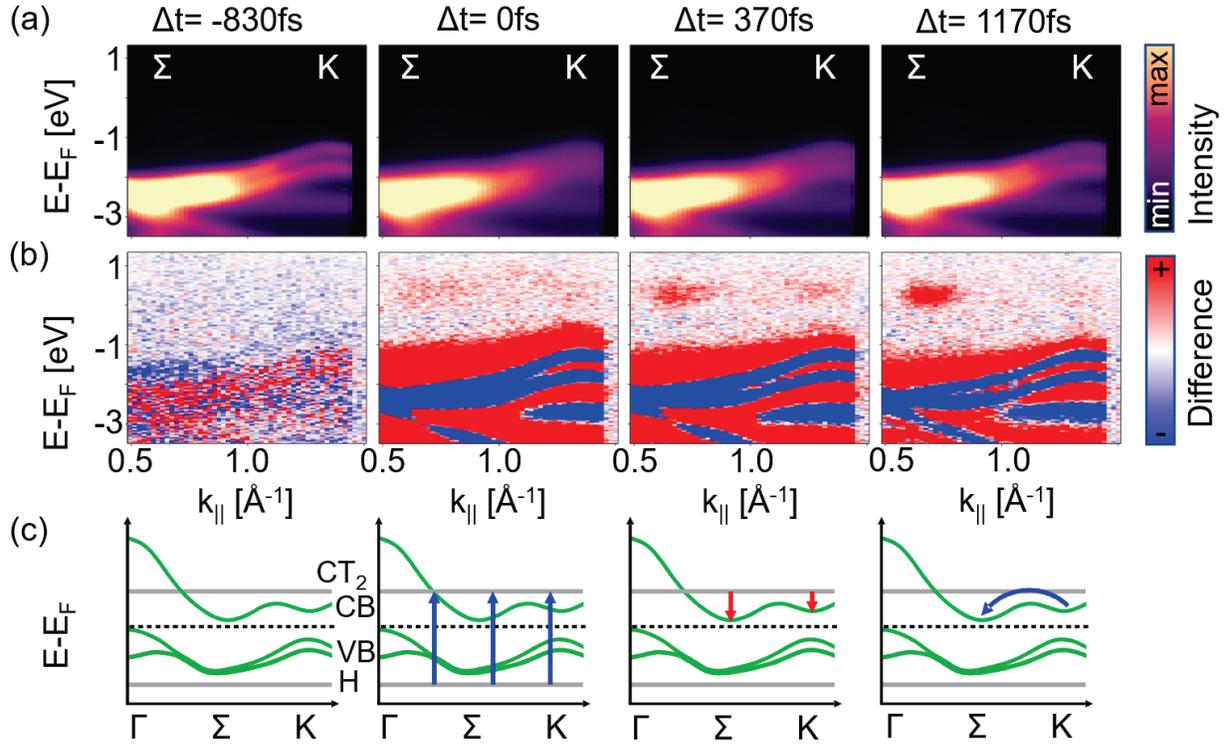

**Fig. S3: Ultrafast dynamics of the $C_{60}$/$WSe_2$ heterostructure.**
(a) Example energy vs. momentum intensity maps at selected time delays after optical excitation of an ultrathin $C_{60}$ film grown on a freshly cleaved $WSe_2$ bulk crystal with 3.2 eV photons (fluence $F = 0.5$ mJ/cm$^2$). The photoemission data were obtained in the normal incidence geometry of the pump and probe beams (p-polarization). (b) Intensity difference maps of the energy-momentum maps shown in (a). The difference maps were calculated by subtracting an intensity map before optical excitation (averaged over several time steps) from the intensity map at a selected time delay. In these difference maps, red areas indicate accumulation of spectral intensity, while blue areas indicate loss of spectral intensity. The carrier dynamics are discussed in the main manuscript and illustrated in the energy level diagram in (c). In contrast to the photoemission data for the bare 2H-$WSe_2$ bulk crystal in Fig. S2, we find a clear population at the K-valley of the $WSe_2$ conduction band due to interlayer charge transfer from the optically excited $C_{60}$ layer. This additional comparison between the raw data of bare and $C_{60}$ covered $WSe_2$ provides further evidence for the ultrafast interlayer charge transfer from $C_{60}$ into the first $WSe_2$ layer, as discussed in the main manuscript.

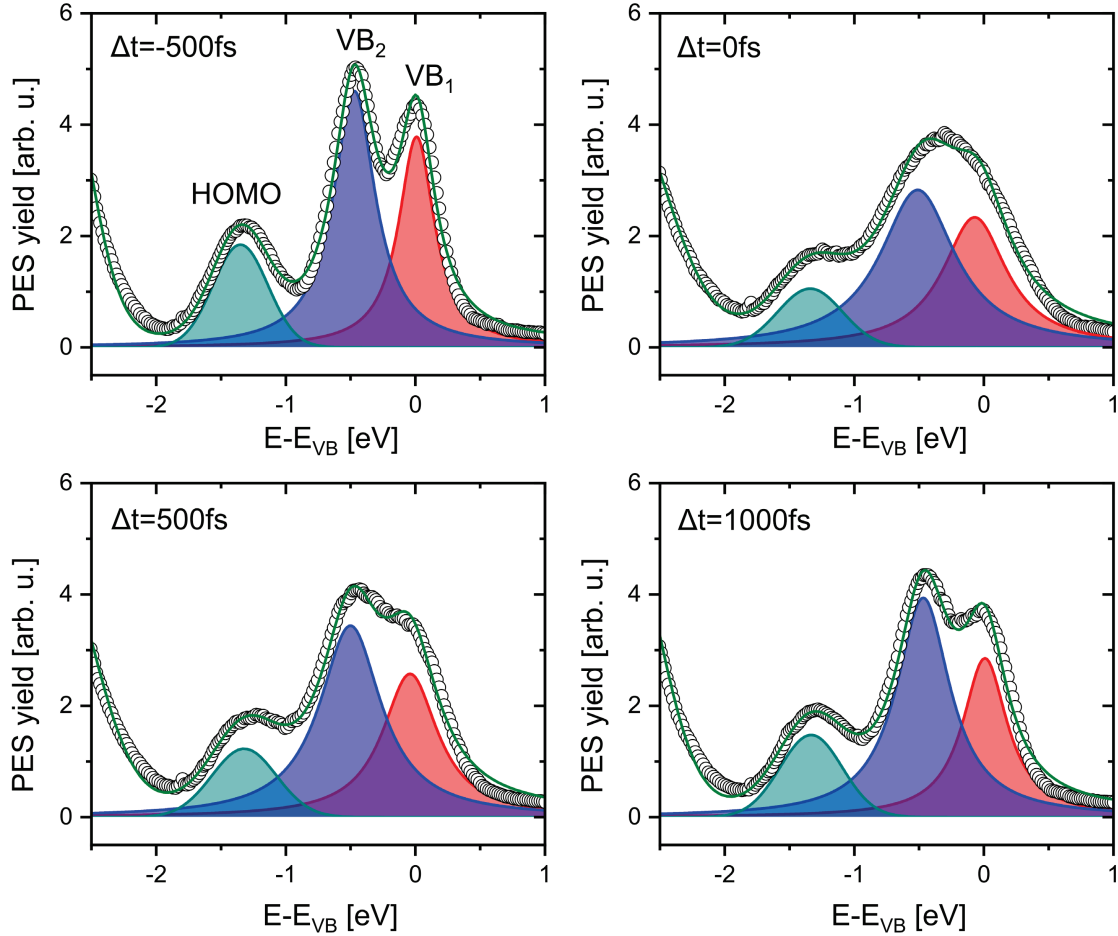

**Fig. S4: Fitting model for the time-and momentum-dependent photoemission data of the $C_{60}$/$WSe_2$ valence band structure.**

The data acquisition of the photoemission yield was performed in the snapshot mode of our photoemission analyzer system using a pass energy of 100 eV. The time-dependent changes of the photoemission signal were obtained by recording a 2D photoemission image (energy vs. emission angle) for each time delay between the 3.2 eV pump and the XUV probe pulse, see Fig. S3. The delay scans were repeated several times to improve the signal-to-noise ratio of the experimental data set. The 1D photoemission spectra of the occupied valence band region were extracted for each time delay from a small angular range (about 2°) around the K-point of the $WSe_2$ Brillouin zone. The analysis of these 1D spectra was performed separately for each spectrum of each time delay scan in two subsequent steps. First, the secondary electron background was subtracted from each spectrum using a Tougaard background [4]. To avoid artifacts in the subsequent fitting procedure, the background subtraction parameters were kept constant for each spectrum of a single delay scan. Example spectra after background subtraction are shown for four selected time delays. A spectral analysis was then performed to quantify the relative changes in peak area A, peak position $E$-$E_{VB}$ and linewidth (FWHM) for each spectroscopic feature, using a dedicated fitting model. The best fit was obtained by the model shown in this figure. The spectroscopic features of the spin-split $WSe_2$ valence bands (VB1 and

VB2) were modeled with Lorentzian functions, the molecular signals (HOMO and HOMO-1) with a Gaussian function.

The best fitting model is plotted as colored curves below the experimental data. Note that the signature of HOMO-1 is outside the energy range of the spectra shown in Fig. S4. The parameters of our fitting model were optimized by fitting the photoemission data at $\Delta t = -500$ fs. This optimized fitting model was propagated to all spectra of the delay scan using a minimum number of fitting constraints. The energy difference between the spin-split valence bands of $WSe_2$ (bands labeled $VB_1$ and $VB_2$) was fixed to $(0.46 \pm 0.03)$ eV, the maximum FWHM of the $C_{60}$ HOMO was set to $(0.16 \pm 0.1)$ eV, and the FWHM of both $WSe_2$ valence bands was limited to be smaller than 0.3 eV. Using these constraints, we were unable to detect any signature of optically induced depletion of the $WSe_2$ valence band after excitation of the $C_{60}/WSe_2$ heterostructure with 3.2 eV pump pulses (see Fig. S5). Therefore, we additionally constrain the range of the Lorentzian functions modeling $VB_1$ and $VB_2$ to $\pm$ 10% of the area obtained for our fit at $\Delta t = -500$ fs. The fitting analysis yields the transient binding energy position and FWHM of the $WSe_2$ valence bands, as well as the binding energy position, FWHM, and area of the $C_{60}$ HOMO. The fitting results of this data analysis procedure are shown in Fig. 2c, and Fig. 3c of the main manuscript and Figs. S5 and S6 of the supplemental material.

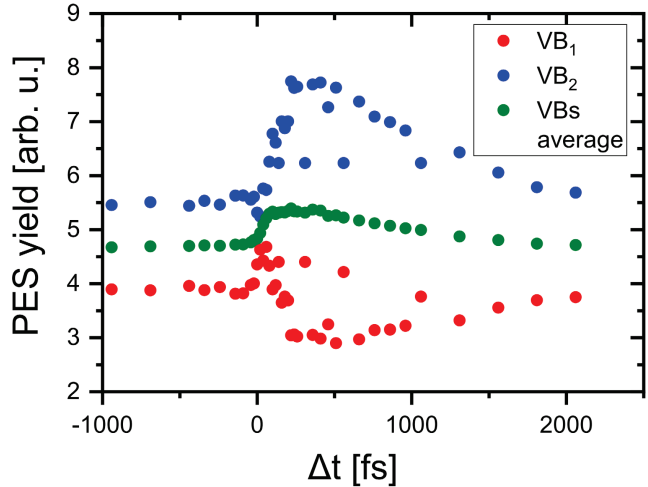

**Fig. S5: Fitting result for the $C_{60}$/$WSe_2$ valence band structure – The area of the $WSe_2$ valence bands.**

To quantify a potential depopulation of the $WSe_2$ valence band after optical excitation with 3.2 eV photons, we plot here the temporal evolution of the area of the Lorentzian functions modeling the valence bands of $WSe_2$. The results of our fitting model suggest a decrease in the intensity of $VB_1$ (red dots) within about 300 fs (i.e., on timescales significantly longer than the duration of the pump pulses) and a substantial increase in the intensity of $VB_2$ (blue dots). In total, this leads to an apparent overall small increase in the intensity of the $WSe_2$ valence bands (green dots). The opposite intensity changes of $VB_1$ and $VB_2$ do not indicate a substantial depopulation of $VB_1$, but rather an imperfect separation of the time-dependent spectral yields of $VB_1$ and $VB_2$. This is not surprising considering that the intensity changes coincide with the time scale of the transient linewidth broadening of the $WSe_2$ valence bands. As shown in Fig. S4, the transient linewidth broadening leads to a merging of the characteristic double peak structure of the spin-split $WSe_2$ valence bands into a single, nearly symmetric spectroscopic feature. Combined with the significantly different excited state dynamics of the bare $WSe_2$ crystal and the $C_{60}$/$WSe_2$ heterostructure, our line-shape analysis of the $C_{60}$/$WSe_2$ valence band structure points to at best small extent of optical excitation of carriers in the bulk $WSe_2$ crystal below the $C_{60}$ layer.

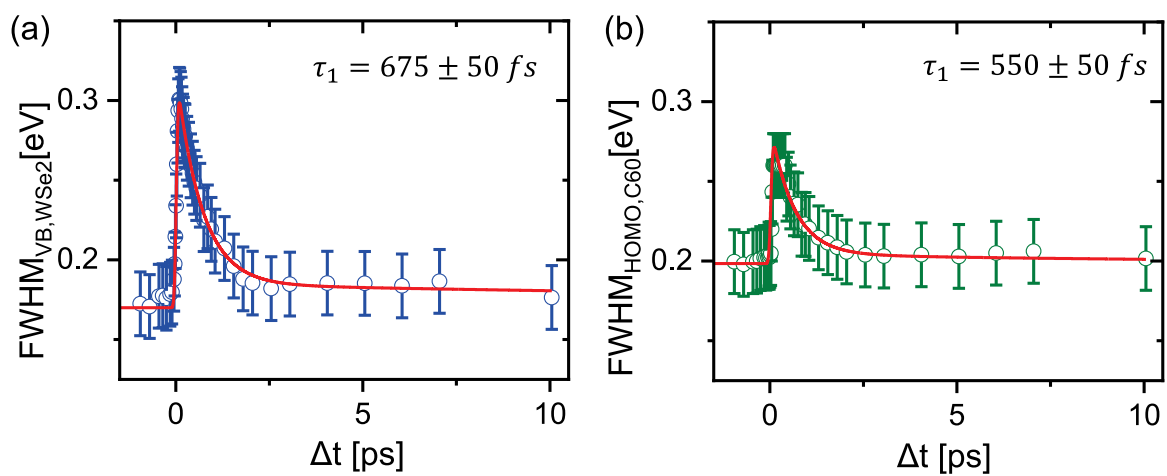

**Fig. S6: Fitting result for the $C_{60}$/WSe$_2$ valence band structure – The transient linewidth broadening of the $C_{60}$/WSe$_2$ valence bands.**

The temporal evolution of the transient linewidth broadening of the WSe$_2$ valence bands (a) and the $C_{60}$ HOMO (b) were determined by the fitting model discussed in the caption of Fig. S4. The temporal evolution of these data was analyzed using a simple exponential fit function with a single decay time $\tau_1$. This exponential function was convolved with a normalized Gaussian with a FWHM of $\Delta t = 70$ fs, and the resulting analytical function was fitted to the experimental data using a least-squares fitting procedure. We find a different temporal evolution of the transient linewidth broadening for the valence states in WSe$_2$ and the $C_{60}$ layer within our experiment uncertainty.

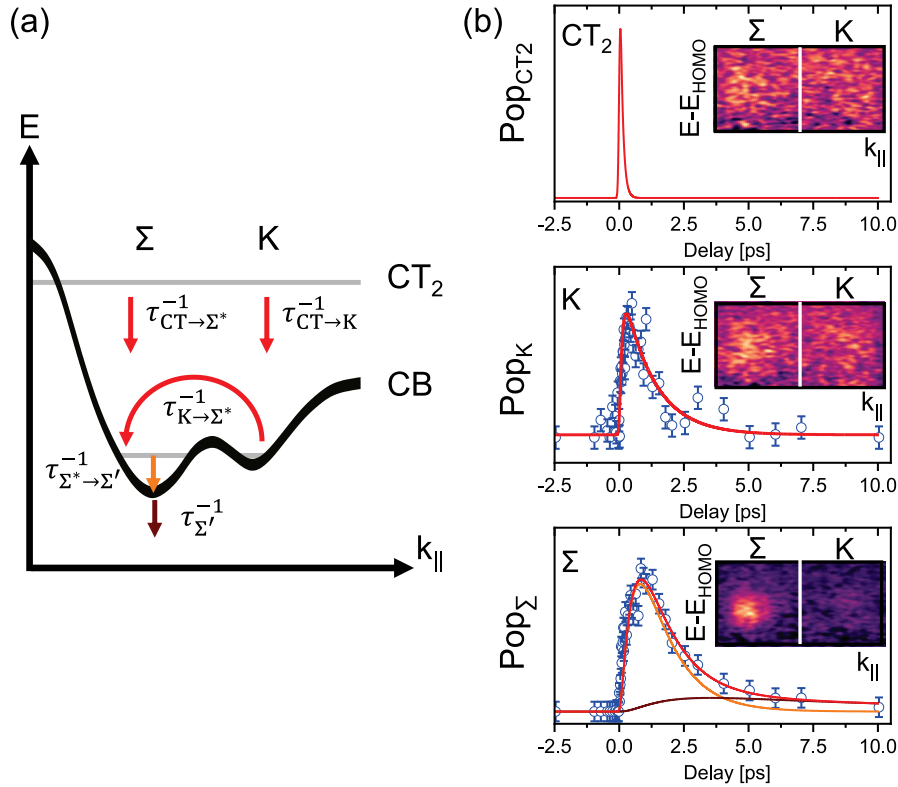

**Fig. S7: Illustration of the rate equation model analysis of the excited state populations.**
(a) Excited state energy level alignment of the $C_{60}$/$WSe_2$ heterostructure. The scattering paths of our rate equation model are indicated by straight and curved arrows together with the corresponding scattering rates considered in the set of rate equations. (b) Comparison of the extracted photoemission yield in the excited states with the simulated time-dependent population traces of our rate equation model. The insets show example energy vs. momentum maps for a selected time during the population of the respective molecular or $WSe_2$ excited state. The vertical white line separates the momentum space region assigned to the excited state population of the $\Sigma$- (left side) and K-valley (right right). The excited population at selected time delay is determined by first numerically integrating the photoemission intensity in the corresponding momentum space region marked in the inset. The resulting excited state spectrum $I_{K/\Sigma}(E,t)$ is then background-corrected with a linear background, and finally fit by a single Gaussian function for each time delay. The intensity of this Gaussian function $A_{K/\Sigma}(t)$ reflects the transient excited state population of the $\Sigma$- and K-valley. Note that despite this dedicated fitting procedure, no clear population signal could be extracted for the $CT_2$ feature due to its broad emission pattern in momentum space as well as its energetic overlap with the $WSe_2$ conduction band.

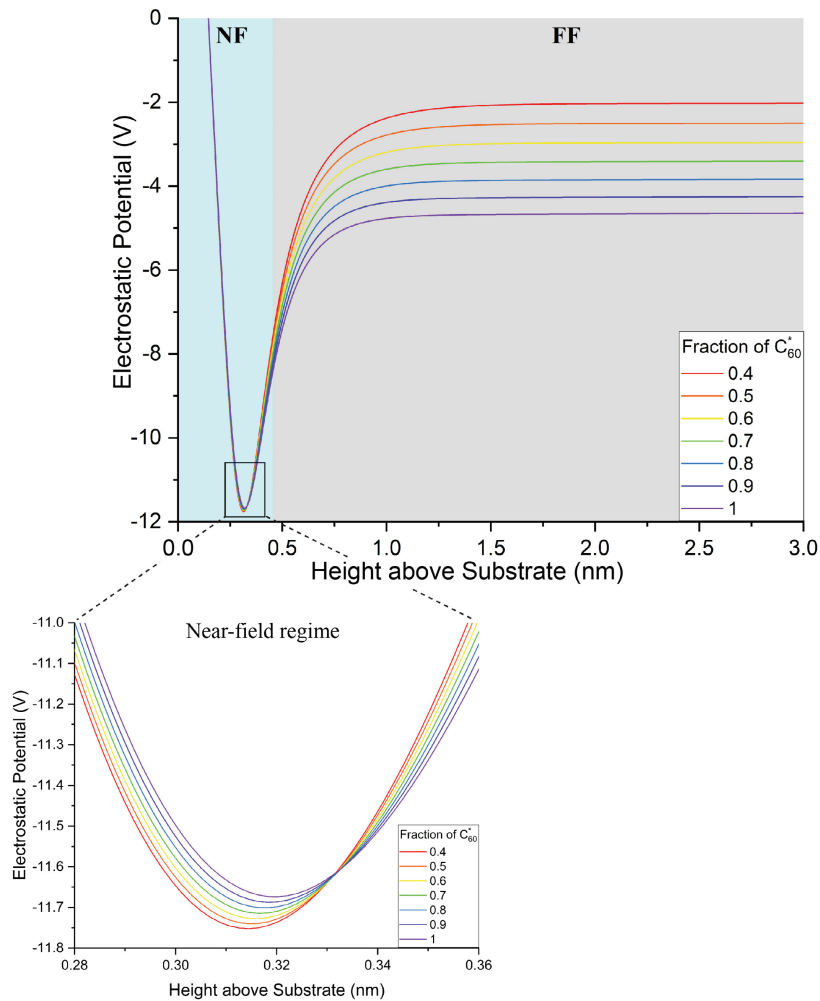

**Fig. S8: Electrostatic potential as a function of height above substrate.**
The blue shaded region is the near-field region (NF), and the grey-shaded region is far-field (FF). The inset shows the switch from NF to FF behavior, where increased excitation densities increase the electrostatic potential within the physical dipole.

## Supplementary Methods

### A. Rate Equation Model:

The temporal evolution of the charge carriers was modeled by solving a rate equation model for the excited state populations $P_i$ of the molecular $CT_2$ state ($i$ = CT), the population at the K-valley of $WSe_2$ ($i$ = K), the high energy population at the $\Sigma$-valley ($i = \Sigma^*$) as well as at the band bottom of the $\Sigma$-valley ($i = \Sigma'$). In our simulations, the initial population is created in the molecular state $CT_2$ using the laser light field $L(t)$, which is modeled as a delta pulse at $t = 0$ fs. The carrier scattering is assumed to be mono-exponential with rates that are given by the inverse of the scattering times $\tau_i$. All possible scattering pathways are illustrated in Fig. S7(a) and the set of rate equations are:

$$\frac{dP_{CT}}{dt} = -\frac{1}{\tau_{CT \to K}} \cdot P_{CT}(t) - \frac{1}{\tau_{CT \to \Sigma^*}} \cdot P_{CT}(t) + L(t)$$

$$\frac{dP_K}{dt} = \frac{1}{\tau_{CT \to K}} \cdot P_{CT}(t) - \frac{1}{\tau_{K \to \Sigma^*}} \cdot P_K(t)$$

$$\frac{dP_{\Sigma^*}}{dt} = \frac{1}{\tau_{CT \to \Sigma^*}} \cdot P_{CT}(t) + \frac{1}{\tau_{K \to \Sigma^*}} \cdot P_K(t) - \frac{1}{\tau_{\Sigma^* \to \Sigma'}} \cdot P_{\Sigma^*}(t)$$

$$\frac{dP_{\Sigma'}}{dt} = \frac{1}{\tau_{\Sigma^* \to \Sigma'}} \cdot P_{\Sigma^*}(t) - \frac{1}{\tau_{\Sigma'}} \cdot P_{\Sigma'}(t).$$

The sum of $P_{\Sigma^*}(t)$ and $P_{\Sigma'}(t)$ corresponds to the total population of electrons at the $\Sigma$-valley of $WSe_2$. However, this separation of the population at the $\Sigma$-valley into two contributions reflects the intravalley carrier scattering reported by Puppin et al. [2] and improves the quality of our data analysis procedure.

To quantify the scattering time, we first analytically solved this set of coupled differential equations and convoluted the resulting time traces with a normalized Gaussian function (FWHM of $\Delta t = 70$ fs) to account for the temporal broadening of our experimental data. Afterwards, we manually fitted the analytical functions for the population of $P_{CT}(t)$, $P_K(T)$, and $P_\Sigma(t) = P_{\Sigma^*}(t) + P_{\Sigma'}(t)$ to the experimental data. For comparison with the measured photoemission intensities $Pop_i$, we used a weighting factor $\alpha_i$: $Pop_i(t) = \alpha_i P_i(t)$. The results are shown in Fig. 2c of the main manuscript and in the supplementary figure S7(b). The most consistent fitting result was obtained for $\tau_{CT\to\Sigma*} \to \infty$, i.e., we neglected a direct charge transfer from the molecular layer into the WSe$_2$ Σ-valley. We would like to emphasize that our results do not allow us to exclude the existence of this charge transfer channel. However, the good agreement between our experimental data and the resulting time-traces of our rate equation model at the K- and Σ-valley suggests that the charge transfer to the Σ-valley is certainly not the dominant interlayer charge transfer channel for the C$_{60}$/WSe$_2$ heterostructure.

## B. Electrostatic Model Simulations

We used a simple electrostatic model to determine the magnitude and direction of the transient electric fields due to charge transfer at the interface of $C_{60}$ and $WSe_2$. Beyond a minor static interfacial dipole, an additional electrostatic potential step emerges as a result of interfacial charge transfer from $C_{60}$ to $WSe_2$. We model the magnitude of this transient electrostatic potential by an array of physical dipole moments $\mu$ per area, $A$, by the Helmholtz equation:

$$\Delta E = \frac{\mu}{\epsilon_0 A} = \frac{q_e d}{\epsilon_0 A}$$

where $\epsilon_0$ is the vacuum permittivity, and $q_e$ the charge separated by distance, $d$. By considering a finite array of dipole moments made up of point charges, we model the electrostatic potential of this interface.

From STM data of $C_{60}$ on $WSe_2$, the reported height of $C_{60}$ on $WSe_2$ is approximately 1 nm, and the radius of $C_{60}$ is 7.1 Å. [5] We assume that the holes, or array of positive charges, are localized at the bottom of the $C_{60}$ cage, while the electrons, or negative charges, are confined to the top-most layer of $WSe_2$, specifically the top-most Se-atom. Due to the van der Waals nature of interlayer binding in $WSe_2$, we do not expect lower layers in $WSe_2$ to participate significantly in the formation of the electrostatic field. The resulting charge separation distance of 2.9 Å establishes electrostatic potentials that qualitatively reproduce the observed maximum transient energy shifts. In what follows, we discuss all aspects of our model and built-in assumptions.

We estimate the excitation densities from the transient depopulation of the HOMO level of $C_{60}$ at key time-steps. Optical excitation of $C_{60}$ with 3.2 eV sub-50 fs pulses resonantly excites electrons from the HOMO level into the LUMO+1* level, leading to the formation of a charge transfer exciton in the $C_{60}$ layer, the $CT_2$ state. [6] The HOMO feature shows an instantaneous intensity reduction of approximately 35%. At the high fluences used in our experiments and based on the photoemission cross section of the $C_{60}$ HOMO bands for s-polarized light [7], we estimate an excitation efficiency of about 80% of all $C_{60}$ in the thin film.

Based on the high density of electron-hole pairs upon interfacial charge transfer, additional effects need to be taken into account for estimating the resulting potentials. Blumenfeld et al. showed previously for thin films of molecules supporting a permanent dipole moment that at high enough dipole densities depolarization effects strongly impact the interfacial electrostatics [8]. Depolarization results from the fact that each dipole moment in the array induces a dipole moment in the surrounding molecules that is aligned in the opposite direction, thus reducing the transient dipole moment. Following their model, the effective transient dipole moment, $\tilde{\mu}_{ind}$ including depolarization effects can thus be expressed as:

$$\tilde{\mu}_{ind} = \frac{\mu_{ind}}{1 + f\tilde{\alpha}_{zz}\rho_{dip}^{3/2}}$$

where $\rho_{dip}$ is the density of transient dipoles, $f$ is a geometric factor capturing the geometric arrangement of dipole moments on the surface and known as the Topping constant [9], and $\tilde{\alpha}_{zz}$ is the *zz*-component of the effective polarizability tensor. We estimate the latter based on the known static polarizability of $C_{60}$ $\alpha_{zz} = 8 \cdot 10^{-29}$ m$^3$ [10], and consider screening effects by both sides of the interface by including their respective dielectric components into the simulation. The static dielectric constant of $C_{60}$ is $\epsilon_{C60} = 4.5$ and the bulk out-of-plane component of the static dielectric constant for WSe$_2$ is $\epsilon_{WSe2} = 7.8$ [10,11]. This yields the *zz*-component of the effective polarizability tensor $\tilde{\alpha}_{zz}$:

$$\tilde{\alpha}_{zz} = \frac{\alpha_{zz}}{4\pi\epsilon_0\epsilon_{avg}}$$

where $\epsilon_{avg}$ is the average dielectric constant of the two materials. By way of comparison, we estimate an induced dipole moment of approximately 13.9 D for a charge separation of 2.9 Å.

The electrostatic potential as a function of height above the surface is shown in Figure S8 for different fractions of excited $C_{60}$ molecules. From the electrostatic simulations, we identify two regimes of the electrostatic surface potential (near- and far-field, highlighted in blue and grey, respectively). At approximately 3.3 Å, we observe an isosbestic-like point where the far-field effects begin to dominate. In the far-field regime, the electrostatic potential above the dipole array decays rapidly, converging to a constant potential step, representing a transient change of the global work function of the system as a result of charge-separation. As expected and given the sign of the dipole moment, an increase in excitation density (charge density) on the surface decreases the electrostatic potential and therefore the work function.

The near-field potential, or the potential "inside" the physical dipole, gives rise to an electrostatic potential whose behavior contrasts with the far-field potential: Due to depolarization, the electrostatic potential decreases with increasing charge densities. The electrostatic potential in the near-field regime explains therefore both signs and magnitude of the Stark shifts observed in the $C_{60}$ HOMO and WSe$_2$ valence band: At lower charge densities, the magnitude of the electrostatic potential close to the dipole array increases.

## Supplementary References